\documentclass[twocolumn]{aastex63}

\usepackage{epsfig}
\usepackage{amsmath}
\usepackage{enumerate}
\usepackage{mathrsfs}
\usepackage{mathtools}
\usepackage{units}
\usepackage{hyperref}
\usepackage{lineno}
\shorttitle{Wind from a Magnetic Monopole Rotator}
\shortauthors{Hu, Beloborodov, \& Chen}
\def\bE{{\mathbf E}}
\def\bB{{\mathbf B}}
\def\RLC{R_{\rm LC}}

\def\bj{\boldsymbol{j}}
\def\bv{\boldsymbol{v}}
\def\br{\boldsymbol{r}}

\def\beq{\begin{equation}}
\def\eeq{\end{equation}}

\def\rhoco{\rho_{\rm co}}
\def\M{{\mathcal M}}
\def\bvD{\boldsymbol{v}_D}
\def\bS{\boldsymbol{S}}
\def\bF{\boldsymbol{F}}

\def\rsat{r_{\rm sat}}

\def\Rout{R_{\rm out}}

\newbox\grsign \setbox\grsign=\hbox{$>$} \newdimen\grdimen \grdimen=\ht\grsign
\newbox\simlessbox \newbox\simgreatbox \newbox\simpropbox
\setbox\simgreatbox=\hbox{\raise.5ex\hbox{$>$}\llap
     {\lower.5ex\hbox{$\sim$}}}\ht1=\grdimen\dp1=0pt
\setbox\simlessbox=\hbox{\raise.5ex\hbox{$<$}\llap
     {\lower.5ex\hbox{$\sim$}}}\ht2=\grdimen\dp2=0pt
\setbox\simpropbox=\hbox{\raise.5ex\hbox{$\propto$}\llap
     {\lower.5ex\hbox{$\sim$}}}\ht2=\grdimen\dp2=0pt

\def\simlt{\mathrel{\copy\simlessbox}}

\definecolor{ao(english)}{rgb}{0.0, 0.5, 0.0}
\begin{document}

\title{Relativistic wind from a magnetic monopole rotator}

\author{Rui Hu}
\affiliation{Physics Department and Columbia Astrophysics Laboratory, Columbia University, 538 West 120th Street New York, NY 10027, USA}

\author{Andrei M. Beloborodov}
\affiliation{Physics Department and Columbia Astrophysics Laboratory, Columbia University, 538 West 120th Street New York, NY 10027, USA}
\affiliation{Max Planck Institute for Astrophysics, Karl-Schwarzschild-Str. 1, D-85741, Garching, Germany}

\author[0000-0002-4738-1168]{Alexander Y. Chen}
% \affil{Department of Astrophysical Sciences, Princeton University, Princeton,
% NJ 08544, USA}
\affil{
JILA, University of Colorado, 440 UCB, Boulder, CO 80309, USA
}

\begin{abstract}
A rotating star with a monopole (or split monopole) magnetic field gives the simplest, prototype model of a rotationally driven stellar wind. Winds from compact objects, in particular neutron stars, carry strong magnetic fields with modest plasma loading, and develop ultra-relativistic speeds. We investigate the relativistic wind launched from a dense, gravitationally bound, atmosphere on the stellar surface. We first  examine the problem analytically and then perform global kinetic plasma simulations. Our results show how the wind acceleration mechanism changes from centrifugal (magnetohydrodynamic) to electrostatic (charge-separated) depending on the parameters of the problem. The two regimes give winds with different angular distributions and different scalings with the magnetization parameter.
\end{abstract}

\keywords{
Rotation powered pulsars (1408);
Plasma astrophysics (1261);
Magnetic fields (994)
% Neutron stars (1108)
}

% %#########################################################################

\section{Introduction}
Rotationally driven outflows are ubiquitous in astrophysics. Compact objects --- neutron stars and accreting black holes --- produce especially powerful winds, because of their strong magnetic fields and fast rotation. They power some of the brightest high-energy sources, such as pulsar wind nebulae and blazars.

A realistic stellar magnetic field may consist of complex multipolar components, however some basic properties of rotationally driven outflows can be studied with the simplest, monopole (or split monopole) field. This simplification was used to construct the pioneering models for non-relativistic winds from rotating stars \citep{weber_angular_1967}, relativistic winds from neutron stars \citep{michel_rotating_1973} and black holes \citep{blandford_electromagnetic_1977}. These models were developed in the framework of ideal magnetohydrodynamics (MHD), where electric field vanishes in the rest frame of the plasma.

The MHD theory has a well defined ultra-relativistic limit of ``force-free electrodynamics'' (FFE), which neglects the plasma inertia and assumes sufficient plasma density to screen the electric field component parallel to the magnetic field, $E_\parallel=0$.  A simple analytical solution in the FFE limit was obtained for the wind from a monopole rotator by \citet{michel_rotating_1973}. However, the MHD and FFE models can fail to predict the correct acceleration of the wind, which may require a more involved consideration of electrostatic acceleration by $E_\parallel\neq 0$ \citep{michel_rotating_1974,fawley_potential_1977}.

In this paper, we study wind launching using first-principles plasma simulations. The plasma is simulated as a large number of individual charged particles moving in the self-consistent electromagnetic field. The field is governed by Maxwell equations, and the evolution of the plasma+field system is followed with the particle-in-cell (PIC) method. Such global kinetic simulations proved useful in recent studies of pulsar magnetospheres (see \cite{Cerutti17} for a review). We focus here on the monopole rotator for three reasons:
\\
(1) As the simplest prototype wind model, the monopole rotator displays some basic features of relativistic plasma winds. Dipole rotators have a different structure inside the light cylinder (part of the magnetosphere is closed), however their winds form in the open field-line zone, which resembles the monopole wind. Our results for the dipole rotator will be presented in a separate paper (Hu \& Beloborodov, in preparation).
\\
(2) We wish to examine the acceleration mechanism of relativistic winds in detail, and elucidate the distinction between the MHD and charge-separated outflows, originally pointed out by \citet{michel_rotating_1974}. We demonstrate that both regimes can co-exist in the magnetosphere of a monopole rotator. This generally occurs for stars with a realistically thin, dense plasma atmosphere. We summarize the efficiency of the wind acceleration in both regimes.
\\
(3) We use the rotating monopole as a test laboratory for our PIC code to test the capabilities of kinetic plasma simulations. We implement a dense, gravitationally bound, atmospheric layer on the stellar surface, which serves as a self-consistent source of plasma particles for the wind. We also push the system to a high magnetization, reducing plasma inertia in the magnetosphere, and study the approach to the force-free electromagnetic configuration.

The paper is organized as follows. Section~\ref{sec:formulation-problem} gives the formulation of the problem, and Section~\ref{sec:force-free-monopole} summarizes the FFE solution. Then, in Section~\ref{sec:wind-accel-mech}, we discuss the MHD and charge-separated regimes of the wind acceleration. Section~\ref{sec:simulation-setup} describes the numerical setup of the simulations, and the numerical results are presented in Section~\ref{sec:simulation-results}.

%####################################################################

\section{Formulation of the problem}
\label{sec:formulation-problem}

The problem of wind launching by a rotating star has three parameters: the star's radius $R_\star$, its angular velocity $\Omega$, and the radial component of the magnetic field at the stellar surface $B_\star$. We will use spherical coordinates $(r,\theta,\phi)$ with the polar axis along the angular velocity vector. Then the rotation velocity of the star is given by 
\beq
 \bv_{\rm rot}(r,\theta)=\boldsymbol{\Omega}\times\br=(0,0,\Omega r\sin\theta).
\eeq
The electromagnetic fields $\bB$ and $\bE$ are fixed below the stellar surface. The magnetic field is monopolar,
\beq
   \bB=(B_r,0,0), \quad B_r=B_\star\frac{r}{R_\star} \qquad (r<R_\star).
\eeq
The star is assumed to be an ideal conductor, and so its electric field must vanish in the co-rotating frame, $\boldsymbol{E}+\boldsymbol{v}_{\rm rot}\times \boldsymbol{B}/c=0$. This gives
\beq
   \bE=(0,E_\theta,0), \quad  E_\theta=-\frac{\Omega r\sin\theta}{c}\,B_r \qquad (r<R_\star).
\eeq
This electric field implies a charge density inside the star,
\beq
   \rho =\frac{\nabla\cdot\bE}{4\pi}=-\frac{\boldsymbol{\Omega}\cdot \bB}{2\pi c} \qquad (r<R_\star),
\eeq
which is called ``co-rotation'' charge density, $\rhoco$, because $\bE$ is induced by the rotation of the star.

The radial component of the magnetic field is continuous at the stellar surface $r=R_\star$. It gives one boundary condition for the magnetosphere, which determines its type (monopole),
\begin{equation}
\label{eq:Bstar}
   B_r(R_\star,\theta)=B_\star.
\end{equation}
Another boundary condition is set by the continuity of the tangential component of the electric field, 
\begin{equation}
\label{eq:Estar}
   E_\theta(R_\star,\theta)=-\frac{\Omega R_\star\sin\theta}{c}\,B_\star, 
   \qquad E_\phi(R_\star,\theta)=0.
\end{equation}
This boundary condition communicates the rotation of the star to the magnetosphere.

Besides providing the boundary conditions for the external electromagnetic field, the stellar surface is also the source of plasma particles. The charged particles, electrons or ions, can be pulled out from the star (or its dense atmospheric layer) if there appears a component of the electric field parallel to the local magnetic field, $E_\parallel=\bE\cdot\bB/B$. Note that $\bE\cdot\bB=0$ below the surface, however it does not have to be continuous, because a surface charge may develop, in particular when the star is modeled as a conductor with no thermal atmosphere. The field components $E_r$, $B_\phi$, $B_\theta$ just above the surface are not known in advance and need to be found by solving the equations of electrodynamics outside the star. This calculation must be done consistently with the dynamics of charged particles pulled out from the star, which create electric current and thus affect the field dynamics.

The self-consistent electrodynamic problem is described by two Maxwell equations,
\begin{eqnarray}
\label{eq:B_Maxwell}
  \frac{\partial\bB}{\partial t} &=& -c\,\nabla\times\bE, \\
  \label{eq:E_Maxwell}
  \frac{\partial\bE}{\partial t} &=& c\,\nabla\times\bB-4\pi\,\bj,
\end{eqnarray} 
and the equation of motion for each magnetospheric particle of charge $\pm e$ and mass $m$,
\begin{equation}
  \frac{d\boldsymbol{p}}{dt}=\pm e\left(\bE+\frac{\boldsymbol{v}\times\bB}{c}\right)+m\boldsymbol{g}.
\end{equation}
Here $\boldsymbol{p}$ is the particle momentum, $\boldsymbol{v}$ is the particle velocity, and $\boldsymbol{g}$ is the gravitational acceleration.\footnote{Here, we limit our consideration to Newtonian gravity. General relativistic corrections can significantly change the wind acceleration near the star.}
Collectively, the motion of plasma particles determines the electric current density $\bj$ that enters one of the Maxwell equations.

A steady electromagnetic configuration may be established around a steadily rotating star. It will obey the steady-state equations, which are obtained from Equations~(\ref{eq:B_Maxwell}) and (\ref{eq:E_Maxwell}) by setting the left-hand side to zero. The wind flowing from the star will pass through the light cylinder and its momentum will approach the final (asymptotic) value beyond a characteristic radius that needs to be calculated. Free escape of the wind is assumed at the outer boundary, i.e. there are no obstacles for its expansion from the star.

Characteristic timescales of the problem are set by the following frequencies:
\\
\begin{itemize}
\item
Angular velocity of the star's rotation, $\Omega$.
\item
Gyro-frequency $\omega_B=eB_\star/mc$ for electrons ($m=m_e$) and ions ($m=m_i$).
\item 
Plasma frequency $\omega_p=(4\pi e^2n/m)^{1/2}$ for electrons ($m=m_e$) and ions ($m=m_i$), where $n=n_e=n_i$ is the plasma density. A characteristic $n$ may be associated with the co-rotation charge density at the stellar surface, $n=|\rhoco|/e=\Omega B_\star|\cos\theta|/2\pi ec$. This gives 
\beq
\label{eq:omega_p}
  \omega_p
  =\left(2\Omega\,\omega_B|\cos\theta|\right)^{1/2}.
\eeq
\end{itemize}
The characteristic spatial scales are $R_\star$, light cylinder radius $\RLC=c/\Omega$, a characteristic Larmor radius $r_{\rm L}=mc^2/eB$, and the plasma skin depth $\lambda_p=c/\omega_p$. The frequencies are normally ordered as follows
\beq
   \Omega\ll\frac{c}{R_\star}\ll\omega_p\ll\omega_B.
\eeq
However, this ordering is violated in the equatorial plane, because $\omega_p\rightarrow 0$ at $\theta\rightarrow\pi/2$.  

The main dimensionless parameter of the problem may be defined as follows \citep{michel_rotating_1973},
\begin{equation}
\label{eq:sigma}
  \sigma \equiv \frac{e B_{\star} R_{\star}^2\, \Omega}{m_e c^3} = \frac{\omega_B}{\Omega} \left( \frac{R_{\star}}{R_{\mathrm{LC}}} \right)^2.
\end{equation}
It characterizes how much magnetic energy is available per particle at the light cylinder, which directly correlates with the terminal Lorentz factor of the wind. The force-free magnetospheric configuration is formed in the limit of $\sigma\rightarrow\infty$.

%####################################################################
\section{Force-free monopole magnetosphere}
\label{sec:force-free-monopole}

In the FFE approximation, the plasma motion is not calculated self-consistently. Instead, the electrodynamic equations for $\bE$ and $\bB$ are closed by the condition
\beq 
\label{eq:FF}
  \rho\bE+\bj\times\bB/c=0, 
\eeq
where $\rho=\nabla\cdot\bE/4\pi$ is the charge density and, in a steady state, $\bj=(c/4\pi)\nabla\times\bB$. The condition~(\ref{eq:FF}) implies conservation of energy and momentum of the electromagnetic field; the plasma is assumed to be so dilute that it cannot accept any significant fraction of the field energy. For a steady wind this implies that the net Poynting flux through a sphere of radius $r$ is constant with $r$. The plasma is assumed to sustain the required $\rho$ and $\bj$, and how this is achieved is irrelevant in the FFE model.

\citet{michel_rotating_1973} found the exact solution to the force-free electromagnetic field around a steadily rotating star with a monopole magnetic field. The solution is
\begin{eqnarray}
  \label{eq:B_FFE}
    \boldsymbol{B} &=&  B_{\star}\frac{R_{\star}^2}{r^2} \,\hat{\boldsymbol{r}} +
                         B_{\phi} \,\hat{\boldsymbol{\phi}}, \quad  B_{\phi} = -\frac{\Omega r \sin\theta}{c} B_r, \\
                         \label{eq:E_FFE}
    \boldsymbol{E} &= & E_{\theta}\,\hat{\boldsymbol{\theta}}, \qquad E_{\theta} = B_{\phi}, 
\end{eqnarray}
Note the presence of $B_\phi\neq 0$ while the poloidal magnetic field remains exactly radial, unchanged from the vacuum monopole solution. The corresponding charge density $\rho=\nabla\cdot\bE/4\pi$ and the electric current density $\bj=(c/4\pi)\nabla\times\bB$ are
\beq
\label{eq:current}
  \rho =  -\frac{\boldsymbol{\Omega}\cdot \bB}{2\pi c}, \qquad
    \bj   =  c \rho \, \hat{\boldsymbol{r}}.
\eeq

Just like inside the rotating star, the external electromagnetic field satisfies $\bE+\boldsymbol{v}_{\rm rot}\times\bB/c=0$, and so the magnetosphere is co-rotating with the star --- the electric field measured in the co-rotating frame vanishes. This condition holds even outside the light cylinder, where no physical frame can co-rotate with the star. In contrast to the stellar interior, the magnetospheric field lines are bent in the azimuthal direction, and $B_\phi\neq 0$ is supported by the radial electric current $\bj$. \citet{michel_rotating_1973} assumed that the required current is carried by a charge-separated flow with speed $c$, consistent with the relation $j=c\rho$ (Equation~\ref{eq:current}). Note that a particle flowing out radially with speed $c$ stays on the same magnetic field line, like a bead on a bent wire co-rotating with the star. The effect of rotation of the wire is offset by its backward bending in $\phi$, so that the particle has $v_\phi=0$.

The Poynting flux of the force-free wind, $\boldsymbol{S}=c\,\bE\times\bB/4\pi$, is given by
\beq
\label{eq:S}
   S_r=c\,\frac{B_\phi^2}{4\pi}, \qquad S_\theta=0, \qquad S_\phi=-c\,\frac{B_rB_\phi}{4\pi}. 
\eeq
Its poloidal component $\bS_{\rm pol}$ is in the radial direction, parallel to $\bj$. Conservation of charge and electromagnetic energy in the steady wind implies $\nabla\cdot\bj=0$ and $\nabla\cdot\bS=\nabla\cdot\bS_{\rm pol}=0$ (using axisymmetry $\partial/\partial\phi=0$). The ratio of the two parallel vectors $\alpha=\bS_{\rm pol}/\bj$ satisfies the identity $\bj\cdot\nabla\alpha=\nabla\cdot (\alpha \bj)=0$, and so $\alpha$ remains constant along the radial streamlines of the electric current.

The  current density $j$ determines a minimum particle flux in the wind $j/e$, which can be used to define a dimensionless quantity, 
\beq
\label{eq:sigma_w}
  \sigma_w=\frac{e\bS_{\rm pol}}{m_ec^2 \bj}
  =\frac{e B_\phi^2}{2 m_ec|\boldsymbol{\Omega}\cdot\bB|} 
  =\sigma\,\frac{\sin^2\theta}{2 |\cos\theta|}, 
\eeq
where $\sigma$ is the dimensional parameter defined in Equation~(\ref{eq:sigma}). The quantity $\sigma_w$ is constant with $r$ and depends only on $\theta$. It represents energy per charge $e$ (in units of $m_ec^2$) flowing in the force-free wind.

\section{Wind Acceleration Mechanism}
\label{sec:wind-accel-mech}

The FFE model gives an accurate description for the electromagnetic field in the limit of $\sigma\rightarrow\infty$. For the plasma, however, it only specifies the drift velocity, 
\beq
 \bv_D=c\,\frac{\bE\times\bB}{B^2}, 
\eeq
which is perpendicular to $\bB$. The model fails to describe the wind acceleration along the magnetic field lines. 

\citet{michel_rotating_1973} considered a charge separated wind, composed of charges of one sign --- electrons if $\rho<0$ and ions if $\rho>0$. Then the relation $j=c\rho$ requires that the charges move with the speed of light everywhere in the magnetosphere outside the star. This picture implies a Lorentz factor $\gamma\rightarrow\infty$ and gives no information on the wind acceleration mechanism or how the actual $\gamma$ scales with $\sigma\rightarrow\infty$. Note also that the charges are extracted from the star, and in reality they cannot start out at $R_\star$ with $v=c$. Their acceleration from the surface requires $\bE\cdot\bB\neq 0$, and so an acceleration model for the charge separated wind must go beyond the FFE approximation, i.e. it should be formulated with a finite $\sigma$ before taking the limit of $\sigma\rightarrow\infty$ \citep{michel_rotating_1974}.

The FFE configuration can also be sustained with a plasma composed of both negative and positive charges, with number densities $n_-$ and $n_+$, respectively. The net charge density
\beq
   \rho=e(n_+-n_-)=\frac{j}{c}
\eeq
requires $n_+\ne n_-$. This mismatch can be relatively small, $|n_+-n_-|\ll n_\pm$, if the charges have a high ``multiplicity'' defined by
\begin{equation}
  \label{eq:M}
 \mathcal{M} \equiv  \frac{F_r}{F_{\min}}, \qquad F_{\min}=\frac{j}{e},
\end{equation}
where $F_r$ is the radial component of the particle flux in the wind with velocity $\bv$,
\beq
   \bF=(n_++n_-)\bv.
\eeq

Below we discuss the two opposite regimes of $\M=1$ (charge separated flow) and $\M\gg 1$ (MHD flow). In both regimes, a relativistic wind is launched by the magnetosphere, however with qualitatively different acceleration mechanisms, resulting in different Lorentz factors. The terminal Lorentz factor of the charge separated wind scales as $\gamma_{\infty}\sim \sigma^{1/2}$ while in the MHD case $\gamma_{\infty}\sim \sigma^{1/3}$, with a different angular dependence of $\gamma_{\infty}(\theta)$.

\newpage 

\subsection{MHD Wind}
\label{sec:mhd-wind}

In the regime of $\M\gg 1$ the abundant charges screen electric fields in the plasma rest frame, enforcing the ideal MHD condition $\bE\cdot\bB=0$, regardless of the value of $\sigma$. Such a wind cannot be accelerated by $E_\parallel\neq 0$.

In the cold approximation (thermal speed much smaller than the fluid speed), the wind motion may be described by the hydrodynamic drift velocity $\bvD=\bE\times\bB/B^2$. One may view the MHD wind acceleration in the lab frame as the result of magnetic stresses (tension and pressure). Alternatively, the wind launching process may be viewed in the co-rotating frame where the centrifugal force drives the plasma outflow. Below we summarize the relativistic MHD wind model known for 5 decades \citep{michel_relativistic_1969,goldreich_stellar_1970}.

For monopole rotators with $\sigma\rightarrow\infty$ the wind velocity $\bv=\bvD$ is found by substituting the FFE solutions for $\bB$ and $\bE$ (Equations~(\ref{eq:B_FFE}) and (\ref{eq:E_FFE})) into $\bvD=c\bE\times\bB/B^2$. This gives
\beq
\label{eq:vD}
   \bv=\bvD=\frac{B_\phi^2}{B^2}\,\hat{\br} - \frac{B_\phi B_r}{B^2}\,\hat{\boldsymbol{\phi}}.
\eeq
Well inside the light cylinder the wind motion is dominated by rotation, $v_\phi\gg v_r$, and far outside the light cylinder the wind becomes radial, $v_r\gg v_\phi$,
\beq
   \frac{v_\phi}{v_r}=-\frac{B_r}{B_\phi}=\frac{r\sin\theta}{\RLC}.
\eeq
The wind Lorentz factor $\gamma=(1-v^2/c^2)^{-1/2}$ is given by
\begin{equation}
  \label{eq:gD}  
   \gamma(r,\theta) = \sqrt{ 1 + \left(  \frac{r \sin\theta}{\RLC} \right)^2}.
\end{equation}
The wind is sub-relativistic inside the light cylinder, $r\sin\theta\ll\RLC$. At large distances from the rotation axis, $r\sin\theta\gg\RLC$, $\gamma$ grows linearly with distance.

As long as the wind particles are supplied only at the inner boundary (the stellar surface), the steady particle flux $\bF=(n_++n_-)\bv$ must satisfy the conservation law $\nabla\cdot\bF=0$. The ratio of the two parallel vectors $\chi=\bS/\bF$ satisfies the identity $n\bv\cdot\nabla\chi=\nabla\cdot (\chi \bF)=0$ (where we have used $\nabla\cdot\bS=0$). Hence $\chi$ remains constant along the streamlines,
\beq
  \chi=\frac{\bS}{\bF} = \mathrm{const}.
\eeq
The ratio of the Poynting flux $\bS$ to the flux of particle energy $\gamma m_ec^2 \bF$ equals $\chi/\gamma m_ec^2$. It decreases with radius as the wind accelerates.

In the FFE limit ($\sigma\rightarrow\infty$), $\gamma$ grows indefinitely with distance (Equation~\ref{eq:gD}). However, for any finite $\sigma$, the FFE model becomes invalid at a sufficiently large radius, and the wind Lorentz factor saturates at a finite value, which depends on $\sigma$. The characteristic saturation radius $\rsat$ cannot exceed the radius where the kinetic power of the wind (found from Equation~(\ref{eq:gD})) becomes comparable with the Poynting flux. The actual $\rsat$ is much smaller. The wind is accelerated radially by the pressure of the toroidal magnetic field, which is communicated by radial fast magnetosonic waves (``fast modes''). The acceleration saturates at the fast magnetosonic radius $r_{\rm sat}$ where the wind speed $v_r$ reaches the fast mode speed $v_f^\prime$ measured in the fluid frame \citep[e.g.][]{kirk_theory_2009}. At $r>r_{\rm sat}$ parts of the flow at different radii are incapable of exchanging the pressure waves and the wind acceleration becomes inefficient.

The speed of radial fast waves $v_f^\prime$ is controlled by the transverse (toroidal) magnetic field and the corresponding magnetization parameter in the fluid frame is 
\beq
   \sigma^\prime\equiv\frac{{B_\phi^\prime}^2}{4\pi \rho_m^\prime c^2}=\frac{m_e}{\M\bar{m}}\,\frac{\sigma_w}{\gamma}.
\eeq
Here $\M\gg 1$ is the particle multiplicity defined in Equation~(\ref{eq:M}), $\sigma_w$ is given in Equation~(\ref{eq:sigma_w}), $B_\phi^\prime\approx B_\phi/\gamma$, $\rho_m^\prime=\bar{m}(n_++n_-)/\gamma$, and $\bar{m}$ is the average particle mass ($\bar{m}=m_e$ for an electron-positron plasma and $\bar{m}\approx m_i/2$ for an electron-ion plasma). The speed of the fast mode and the corresponding Lorentz factor $\gamma_f^{\prime}=(1-{v_f^\prime}^2/c^2)^{-1/2}$ are given by\footnote{This expression holds for a cold MHD fluid, which is usually a good approximation for winds. When thermal energy is taken into account, the fast magnetosonic speed changes to $v_f^\prime/c=[(\hat{\alpha}-1)w+\sigma^\prime]/(1+w+\sigma^\prime)$, where $w$ is the dimensionless enthalpy and $\hat{\alpha}$ is the adiabatic index of the fluid \citep{beloborodov_sub-photospheric_2017}.} 
\beq
\label{eq:vf}
   \frac{v_f^\prime}{c}=\left(\frac{\sigma^\prime}{1+\sigma^\prime}\right)^{1/2}, \qquad 
   \gamma_f^\prime={\sigma^\prime}^{1/2}.
\eeq
We here focus on winds with $\sigma_w\gg 1$, for which $\rsat\gg \RLC$. Far outside the light cylinder, the magnetic field is dominated by its toroidal component, $B\approx |B_\phi|$ and the wind velocity $\bv=\bvD$ is nearly radial, so that $v_r\approx v$. The critical fast magnetosonic point may be evaluated from the condition 
\beq
\label{eq:point}
  \gamma\approx \gamma_f^\prime
  \approx \left(\frac{m_e}{\M\bar{m}}\,\frac{\sigma_w}{\gamma}\right)^{1/2}, 
\eeq
where $\gamma$ is the Lorentz factor of the force-free wind (Equation~\ref{eq:gD}). Solving Equation~(\ref{eq:point}) for $\gamma$, one finds the saturation Lorentz factor of the wind,
\beq
\label{eq:g_infty}
  \gamma_\infty(\theta)\approx \left(\frac{m_e\sigma_w}{\bar{m} \M}\right)^{1/3}
  = \left(\frac{m_e\sigma}{\bar{m}\M}\,\frac{\sin^2\theta}{2 |\cos\theta|}\right)^{1/3},
\eeq
where Equation~(\ref{eq:sigma_w}) for $\sigma_w$ has been used. The factor of $\sin^2\theta$ in Equation~(\ref{eq:g_infty}) comes from $B_\phi^2$ (the pressure of the toroidal magnetic field is responsible for the wind acceleration); $\cos\theta$ appears in the denumerator (and vanishes at $\theta=\pi/2$) because the particle multiplicity $\M$ was defined in terms of $j$, which vanishes in the equatorial plane. The product $\M\cos\theta$ stays finite at $\theta=\pi/2$.

The well-known result $\gamma_\infty\sim\sigma^{1/3}$ was obtained by \citet{michel_relativistic_1969} and \citet{goldreich_stellar_1970}. The saturation radius $r_{\rm sat}$ may be estimated using the FFE approximation $\gamma\approx r\sin\theta/\RLC$ at radii $r\simlt r_{\rm sat}$ (Equation~(\ref{eq:gD})), which gives
\beq
  r_{\rm sat}\sim \frac{\gamma_{\infty}\RLC}{\sin\theta}.
\eeq

\subsection{Surface atmosphere at the base of the wind}

The star may have a gravitationally bound atmosphere with some temperature $T\neq 0$, which supplies particles to the wind. In an ideal MHD wind with $E_\parallel=0$, the escaping particle flux $F$ is controlled by the effective potential,
\begin{equation}
\label{eq:V}
  V(r,\theta) = - \frac{g_0 R_{\star}^2}{r} - \frac{1}{2} \Omega^2 r^2 \sin^2\theta.
\end{equation}
It takes into account the centrifugal effect. The strong magnetic field near the star implies that the particles are ``magnetized'', so that they move along the magnetic field lines like beads on wires. The field lines (with footprints  frozen in the conducting star) co-rotate with the star. Therefore, the plasma atmosphere co-rotates and experiences a smaller effective gravitational acceleration, reduced by the centrifugal effect. The effective potential has a maximum
\begin{equation}
    \label{eq:Vmax}
    V_{\max}(\theta) = -\frac{3}{2} g_0R_{\star} \left(\frac{\Omega^2R_{\star}}{g_0} \sin^2\theta \right)^{1/3}.
\end{equation}

An isothermal hydrostatic atmosphere is described by its temperature $T=\mathrm{const}$ and density $n_0$ at $r=R_\star$. The atmospheric particles at $r>R_\star$ form a Boltzmann distribution with density 
\beq
\label{eq:n}
    n(r,\theta) = n_0 \exp \left\{-\frac{m[V(r,\theta)-V_\star(\theta)]}{kT}\right\},
\eeq
where $V_\star(\theta)=V(R_\star,\theta)$. The hydrostatic atmosphere extends up to the escape radius $r_{\rm esc}(\theta)$ where the remaining potential barrier $V_{\max}-V$ becomes comparable to the particle kinetic energy $mv_{\rm th}^2/2\sim kT$, 
\begin{equation}
    \label{eq:escape}
    \frac{v_{\rm th}^2}{2} = V_{\max}(\theta)-V(r_{\rm esc}, \theta).
\end{equation}
This is a cubic equation for $r_{\rm esc}$ with the following solution,
\begin{equation}
    \label{eq:escape-radius}
    \frac{r_{\rm esc}(\theta)}{R_\star} = \frac{6g_0R_\star}{v_{\rm th}^2} (\zeta-\zeta^{1/3}) \cos \left( \frac{\pi}{3} + \frac{1}{3}\arccos \frac{1}{\zeta}  \right),
\end{equation}
where
\beq
   \zeta^{2/3} = 1 -  \frac{v_{\mathrm{th}}^2}{2V_{\max}(\theta)}.
\eeq
The limit of $\theta\rightarrow 0$ gives $r_{\rm esc} \rightarrow 2g_0R_{\star}^2/v_{\mathrm{th}}^2$. The effective potential has no centrifugal part on the rotation axis and so $r_{\rm esc}(\theta=0)$ is independent of $\Omega$.

The escaping particle flux may be estimated as 
\beq
\label{eq:F}
   F(\theta)\sim v_{\rm th}\, n(r_{\rm esc})
 \sim v_{\rm th}\, n_0 \exp \left[\frac{V_\star(\theta)-V_{\max}(\theta)}{kT}\right].
\eeq
It determines the plasma loading of the wind in the ideal MHD regime. This regime is satisfied if
\beq
   F\gg \frac{j}{e}  \qquad {\rm (MHD~regime)}.
\eeq

\subsection{Charge-Separated Wind}
\label{sec:charge-separ-wind}

Next, we consider the opposite regime where the centrifugally assisted outflow from the atmosphere is insufficient to feed the electric current with $\M=eF/j\gg 1$. Then, a parallel electric field $E_\parallel$ is induced to sustain the current. The $E_\parallel$ extracts from the atmosphere charges of one sign, forming  a charge separated wind.

An exactly force-free electromagnetic configuration is deficient because it has $E_\parallel=0$ and at the same time implies that the charge-separated wind has speed $v=c$, because $j=c\rho$. A physical wind solution must start at the stellar surface with a much smaller velocity. In particular, observed neutron stars typically have surface temperatures of $0.1-1$~keV, and so are capable of supplying only non-relativistic particles. Their characteristic atmospheric scale-height is tiny, $h\sim kT/mg\simlt 10\,$cm. Acceleration of the charge-separated wind extracted from the star (or from the upper layers of its atmosphere) must be accomplished by a non-zero $E_\parallel$. The electromagnetic field can still be very close to the FFE solution, because a small $E_\parallel\ll B$ is sufficient to accelerate particles to ultra-relativistic speeds.

In a steady state, $\partial\bB/\partial t=-c\nabla\times\bE=0$ implies that the accelerating electric field can be described by an electrostatic potential, $\bE=-\nabla \Phi$. The electrostatic acceleration of a charge-separated wind was first analyzed by \citet{michel_rotating_1974} and later by \citet{fawley_potential_1977}, with somewhat different results. The acceleration effect is associated with a small correction $\delta \bE$ to the co-rotation electric field $\bE_{\rm co}=-\bv_{\rm rot}\times\bB/c$, and one can define the accelerating electrostatic potential $\delta\Phi$, 
\beq
\label{eq:E}
   \bE=-\nabla(\Phi_{\rm co}+\delta\Phi),
\eeq
where
\beq
    \Phi_{\rm co}=-\frac{\Omega B_\star R_\star^2}{c}\,\cos\theta + \mathrm{const}
\eeq 
gives the co-rotating electric field $\bE_{\rm co}$ of the FFE configuration; $\Phi_{\rm co}$ defines $\bE_{\rm co}\perp\bB$ and thus cannot accelerate particles along the magnetic field lines. The wind acceleration away from the star is determined by $\delta\Phi(r,\theta)$ at $r>R_\star$. It satisfies the equation,
\beq
   \nabla^2 \delta\Phi = -4 \pi \left(\rho -  \rhoco\right),
\eeq
where $\rhoco$ is given in Equation~(\ref{eq:current}), and $\rho$ is the actual charge density in the self-consistent acceleration solution. Note that $\delta\Phi=0$ inside the star, and continuity of the electrostatic potential at the stellar surface gives the boundary condition
\beq
   \delta\Phi(R_\star,\theta)=0.
\eeq 

The magnetized particles move along the magnetic field lines, whose shape and rotation are almost unchanged from the FFE solution as long as $\sigma\gg 1$. Therefore, the azimuthal speed of the wind is related to its radial speed by
\beq
\label{eq:vphi}
   v_\phi = v_{\rm rot}+\frac{B_\phi}{B_r}\,v_r = v_{\rm rot} \left(1-\frac{v_r}{c}\right).
\eeq
It can vanish only for particles moving with the speed of light: $v_r=c$ and $v_\phi=0$. Since the wind starts out at the stellar surface with $v_r\ll c$, the plasma motion must have a large azimuthal component (similar to the MHD solution in Section~\ref{sec:mhd-wind}).

For the rest of this section we take the limit of $T\rightarrow 0$. This corresponds to the initial $v_r(R_\star)=0$ and the zero thickness of the atmosphere, so the particle acceleration by $E_\parallel$ starts at $r=R_\star$. The speed $v$ of the charge-separated wind is set by the energy gain in the potential $\delta\Phi$. This also implies that the wind is cold, i.e. there is only bulk speed, with no random velocity component. The wind Lorentz factor is given by 
\beq
  \gamma(r,\theta)=\gamma(R_\star,\theta)+\frac{e |\delta\Phi|}{mc^2}, \qquad \gamma=\left(1-\frac{v_\phi^2}{c^2}-\frac{v_r^2}{c^2}\right)^{-1/2}.
\eeq

\citet{michel_rotating_1974} proposed an approximate formula for the asymptotic Lorentz factor of the wind at $r\gg R_\star$,
\begin{equation}
    \label{eq:Michel}
    \gamma_{\infty}(\theta) = 1 + \sigma^{1/2} \cos\theta.
\end{equation}
His derivation of this result assumed $\delta\Phi\propto \Phi_{\rm co}\propto \cos\theta$ and also assumed that the electric current $j$ is unchanged from that in the force-free configuration. \citet{fawley_potential_1977} relaxed these assumptions and obtained the following result,
\begin{equation}
    \label{eq:Fawley}
    \gamma_{\infty} = 1 + \frac{\sigma^{1/2}}{\sqrt{2}} \sum\limits_{l=0}^{\infty}\,\frac{(-1)^l(2l+1.5)}{(2l+0.5)(2l+2.5)(l+1)}\, P_{2l+1}(\cos\theta),
\end{equation}
where $P_{2l+1}(\cos\theta)$ are the Legendre polynomials.

Before comparing these previous results with our numerical simulations, let us make simple estimates. In particular, it is useful to see how the scaling $\gamma_\infty\propto \sigma^{1/2}$ appears in the limit of $\sigma\gg 1$.

In the nearly force-free regime $\sigma\gg 1$, the deviations of the electromagnetic fields $\delta\bE$ and $\delta\bB$ from the force-free solution (Equations~(\ref{eq:B_FFE}) and (\ref{eq:E_FFE})) are small. In a leading approximation, the charge-separated wind is accelerated along the unperturbed magnetic field lines. In particular, the flow projection onto the poloidal plane follows the poloidal (radial) magnetic field lines, and creates the poloidal electric current that sustains $B_\phi$,
\beq
  j_r=v_r\rho.
\eeq
Near the stellar surface, where $v_r\ll c$, the charge density $\rho$ greatly exceeds $\rhoco$, which results in a steeply increasing $\delta\bE$ according to $\nabla\cdot\delta\bE=4\pi(\rho-\rhoco)$. As a result, the wind accelerates to nearly $c$ on a short scale $r-R_\star\ll R_\star$, so that the conditions $v_r\ll c$ and $\rho/\rhoco\gg 1$ hold only in a thin layer near the stellar surface. 

At radii $r-R_\star\ll R_\star$ one can approximate $\nabla\approx (\partial_r,0,0)$, and the accelerating potential is described by the equation,
\beq
   \frac{d^2}{dr^2} \delta \Phi=-4\pi\rhoco\left(\frac{c}{v_r}-1\right) \qquad (r-R_\star\ll R_\star),
\eeq
where we used the approximation $j_r\approx c\rhoco$ in the leading order, which corresponds to the FFE configuration of $B_\phi$. The corresponding equation for $\gamma=e|\delta\Phi|/mc^2$ reads
\beq
\label{eq:dg}
   \frac{d^2\gamma}{dr^2}=\frac{1}{\lambda_p^2}\left(\frac{c}{v_r} -1 \right) 
   \qquad (r-R_\star\ll R_\star),
\eeq
where $\lambda_p=c/\omega_p$ and $\omega_p=(2\omega_B\Omega\cos\theta)^{1/2}$ (Equation~\ref{eq:omega_p}). Combining Equations~(\ref{eq:dg}) and (\ref{eq:vphi}) one can solve for $v_r(r)$ and then also find $\delta\Phi(r)$.

One can see from Equation~(\ref{eq:dg}) that the wind enters the ultra-relativistic regime $\gamma\gg 1$ above the characteristic altitude $r-R_\star\sim\lambda_p$. The column density of charge created by $\rho/\rhoco>1$ in the sub-relativistic zone $r<R_\star+\lambda_p$,  is comparable to $\rhoco\lambda_p$. At larger altitudes the wind  continues to accelerate linearly in the approximately constant electric field 
\beq
\label{eq:Er}
   E_r\sim 4\pi \lambda_p\rhoco\approx \mathrm{const} \qquad (\lambda_p\ll r-R_\star\ll R_\star),
\eeq
which is created by the column charge density $\rhoco\lambda_p$ of thickness $\sim\lambda_p$ on the surface of the star. The linear acceleration,
\beq
\label{eq:gam}
   \gamma\sim \frac{r-R_\star}{\lambda_p}   \qquad (\lambda_p\ll r-R_\star\ll R_\star),
\eeq
ends where the approximation $r-R_\star\ll R_\star$ breaks, i.e. at altitudes comparable to $R_\star$. This gives a simple estimate for the asymptotic Lorentz factor reached by the wind at $r\gg R_\star$,
\beq
\label{eq:gamma_infty}
   \gamma_\infty\sim \frac{eE_r R_\star}{mc^2}\sim \frac{R_\star}{\lambda_p}\sim \sqrt{\sigma}\,\cos\theta.
\eeq
Note that this estimate of $\gamma_\infty$ neglected the deviation of $j_r$ from $c\rhoco$, i.e. neglected the deviation of $B_\phi$ from the FFE configuration. This deviation gives an additional term in Equation~(\ref{eq:gam}),
\beq
\label{eq:gam1}
   \gamma\sim \frac{r-R_\star}{\lambda_p} + \frac{(r-R_\star)^2}{2\lambda_p^2}\,\left(\frac{j_r}{c\rhoco}-1\right).
\eeq
The estimate~(\ref{eq:gamma_infty}) for $\gamma_\infty$ assumes $|j_r/c\rhoco-1|\simlt \gamma_\infty \lambda_p^2/R_\star^2\sim \gamma_\infty^{-1}$. This condition is (marginally) satisfied, as may be seen from the following consideration.

The deviation from the FFE configuration changes the Poynting flux from the star $\bS$, and part of the Poynting flux is spent to accelerate the wind through the work of the electric force: $\nabla \cdot\bS=-\bE\cdot\bj$. Assuming that the change of the Poynting flux from the FFE model (Equation~\ref{eq:S}) is comparable to the kinetic power of the wind, we can estimate
\beq
    \delta S_r\sim \frac{\gamma_\infty mc^2 j_r}{e} \qquad (r\gg R_\star).
\eeq
This corresponds to a change in $B_\phi$,
\beq
  \frac{\delta B_\phi}{B_\phi} \sim \frac{\delta S_r}{S_r}
  \sim \frac{\gamma_\infty}{\sigma_w} \sim \sigma^{-1/2}.
\eeq
Evaluating the numerical coefficient in this estimate and its dependence on $\theta$ is more difficult, because the electromagnetic energy is re-distributed in latitude before escaping to infinity. Indeed, $E_r\neq 0$ introduces a Poynting flux in the $\theta$-direction, $S_\theta=-cE_r B_\phi/4\pi\sim -\lambda_p \rhoco B_\phi$. It creates an energy flow in the direction from the equatorial plane to the polar axis ($S_\theta\propto -\cos\theta\sin\theta$). The divergence of $\bS$, which feeds the wind acceleration, has an important contribution from $S_\theta$. In particular, on the polar axis $S_r=0$ and $\partial_r S_r=0$. The radial Poynting flux on the axis cannot power the wind acceleration to $\gamma_\infty\sim\sigma^{1/2}$. Thus, the wind near the axis is powered by $\nabla\cdot\bS\approx r^{-1}\partial_\theta S_\theta$.

 %####################################################################
\section{Simulation Setup}
\label{sec:simulation-setup}

We have developed a relativistic particle-in-cell (PIC) code \texttt{Pigeon}, a descendent of \texttt{Aperture} \citep{chen_electrodynamics_2014}.\footnote{\texttt{Pigeon} is available at https://github.com/hoorayphyer/Pigeon.}  We use a spherical coordinate grid spaced logarithmically in the radial direction. This allows us to expand the simulation domain and follow the wind dynamics well beyond the light cylinder $\RLC=c/\Omega$. In the simulation, $\RLC=4R_\star$ and the outer boundary is at $\Rout\approx 30R_\star$. The outer boundary allows free escape of the particles and the electromagnetic field. This is accomplished by implementing a smooth damping layer, with negligible wave reflection. A similar damping layer was used in the pulsar simulations of \citet{chen_electrodynamics_2014}. The inner boundary of the computational domain $r=R_\star$ is the surface of a conducting solid crust rotating with angular velocity $\Omega$. The inner boundary conditions for the electromagnetic field are given by Equations~(\ref{eq:Bstar}) and (\ref{eq:Estar}).

The simulation starts with a non-rotating star, so the initial field configuration is spherically symmetric. Then we gradually spin up the star to $\Omega=(1/4)(c/R_\star)$ and let the system relax to a steady state. The entire evolution and the final state are symmetric about the rotation axis, and our simulations are axisymmetric. We use the log-spherical grid in $r,\theta$ with the size of $1024\times 1024$.

Just above the stellar surface, the simulation maintains a dense plasma atmosphere, which serves as the source of particles for the wind. The atmosphere is bound by the gravitational field of the star,
\beq
  g(r) = g_0\,\frac{R_{\star}^2}{r^2}.
\eeq
The dense surface layer is maintained by constantly injecting warm (Maxwellian) particles from the surface, and the star absorbs particles that fall back to the surface. As a result a steady Boltzmann distribution is established, with the plasma density exponentially decreasing with altitude $r-R_\star$, 
\beq
   n(r)\approx n_0\exp\left(-\frac{r-R_\star}{h}\right) \qquad  (r-R_\star\ll R_\star). 
\eeq
Neglecting the centrifugal acceleration, one can estimate a characteristic hydrostatic scale-height as $h\approx v_{\rm th}^2/2g_0$, where $v_{\rm th}$ is the average thermal speed of injected particles. In the simulation, particles are injected with radial velocities in the co-rotating frame of the star, i.e. with a fixed azimuthal speed $v_{\phi}=v_{\rm rot}=\Omega r\sin\theta$ in the lab frame. 

All our simulations have $v_{\rm th}=0.2c$ in the corotating frame, and the atmosphere scale-height and density are controlled by two other parameters: $g_0$ and the particle injection rate $\dot{N}$. We adjust the injection rate per unit area 
\beq
  F_0 = \frac{d\dot{N}}{dA}=\M_{\rm atm}\,\frac{j}{e}, 
\eeq
so that $\M_{\rm atm}\sim 10$. This is more than sufficient to sustain the electric current. Then, the magnetospheric electric field $E_\parallel=\bE\cdot\bB/B$ (if induced to lift atmospheric particles and feed $\bj$ in the magnetosphere) is screened from the stellar surface by the atmosphere. 

We use $g_0$ as a knob that controls $h$. It determines the altitude at which the atmospheric density $n(r)$ decreases below $j/c$ and $E_\parallel$ develops to sustain $j$ \citep{beloborodov_corona_2007}. Below we perform simulations representing two opposite regimes:
\medskip
\\
{\bf Model~I: MHD regime.} When $g_0$ is sufficiently small (and so $h$ is large), the atmosphere begins to overflow the gravitation potential barrier and escape into the wind before $n(r)$ falls below $j/ec$. In this case, an MHD wind is formed, carrying both negative and positive charges. In the simulation presented below, we model the MHD regime by simply setting $g_0=0$, so that the atmosphere becomes unbound, filling the entire magnetosphere with $n\gg j/ec$.
\medskip
\\
{\bf Model~II: Charge-separated regime.}
When we increase $g_0$ (and thus reduce $h$), ``charge starvation'' occurs, and $E_\parallel$ develops at small altitudes above the atmosphere. It lifts charges of only one sign into the magnetosphere, forming a charge separated wind. In the simulation shown below, this regime is modeled with $g_0 = 0.5 c^2/R_{\star}$, which gives $h \approx 0.04 R_{\star}$.

The positive and negative charges in our simulations will have the same mass $m$. The generalization to unequal masses is straightforward and does not qualitatively change the wind picture. 

The key parameter of the problem is $\sigma$ (Equation~\ref{eq:sigma}). We choose $\sigma=2.5\times 10^3$ for both simulations. The large $\sigma$ implies that the magnetosphere will be very close to the FFE configuration, in both charge separated and MHD regimes.

The parameter $\sigma$ determines the characteristic density $n_{\rm co}=|\rho_{\rm co}|/e=\Omega B_\star/2\pi c e$ evaluated at the pole of the star. This density defines a characteristic plasma frequency $\omega_p=(4\pi e^2 n_{\rm co}/m)^{1/2}$, which satisfies the relation,
\beq
   \frac{c}{\omega_p R_\star}=\frac{1}{\sqrt{2\sigma}}\approx 1.4\times 10^{-2}.
\eeq

%####################################################################
\section{Simulation Results}
\label{sec:simulation-results}

Below we present the results after a steady wind is established. The steady state is approached in the entire simulation box on the light crossing timescale $\sim \Rout/c$. 

% %%%%%%%%%%% FIGURE %%%%%%%%%%%%%%%%%%
\begin{figure*}[t!]
  \begin{tabular}{cc}    
    \includegraphics[width=0.46\textwidth]{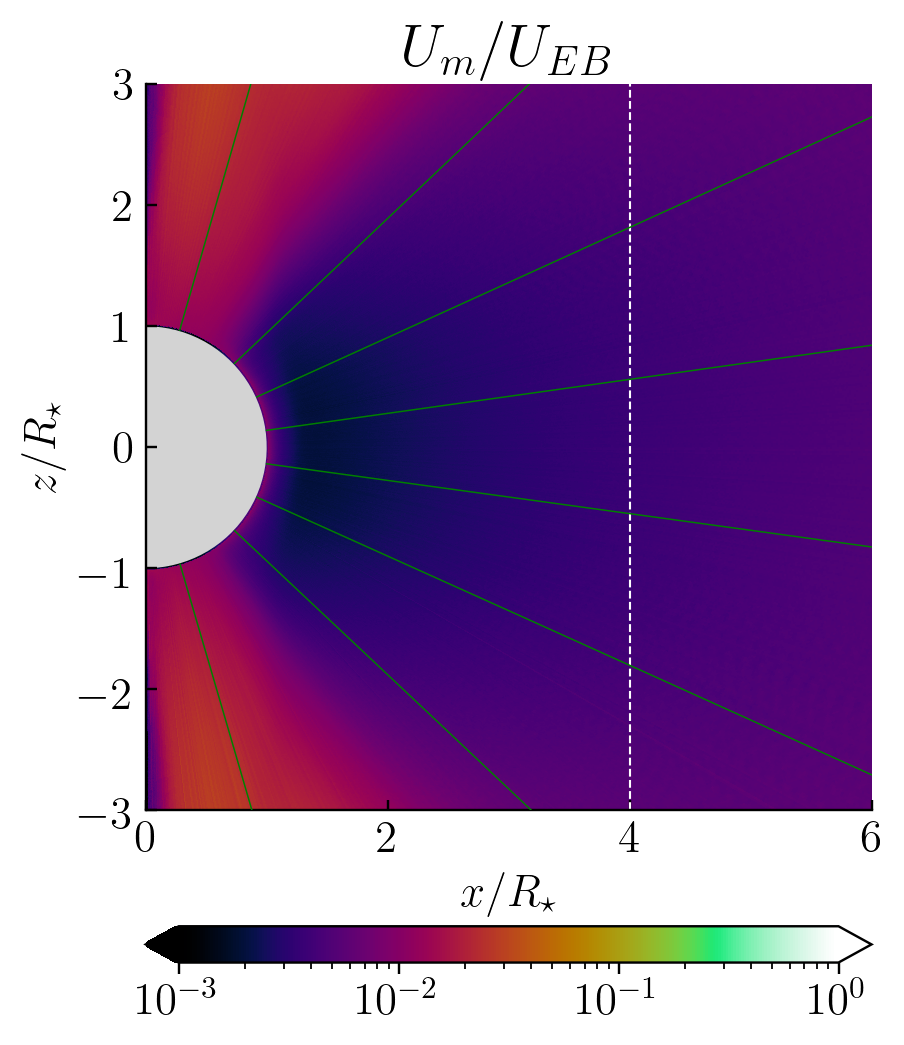} 
    \includegraphics[width=0.46\textwidth]{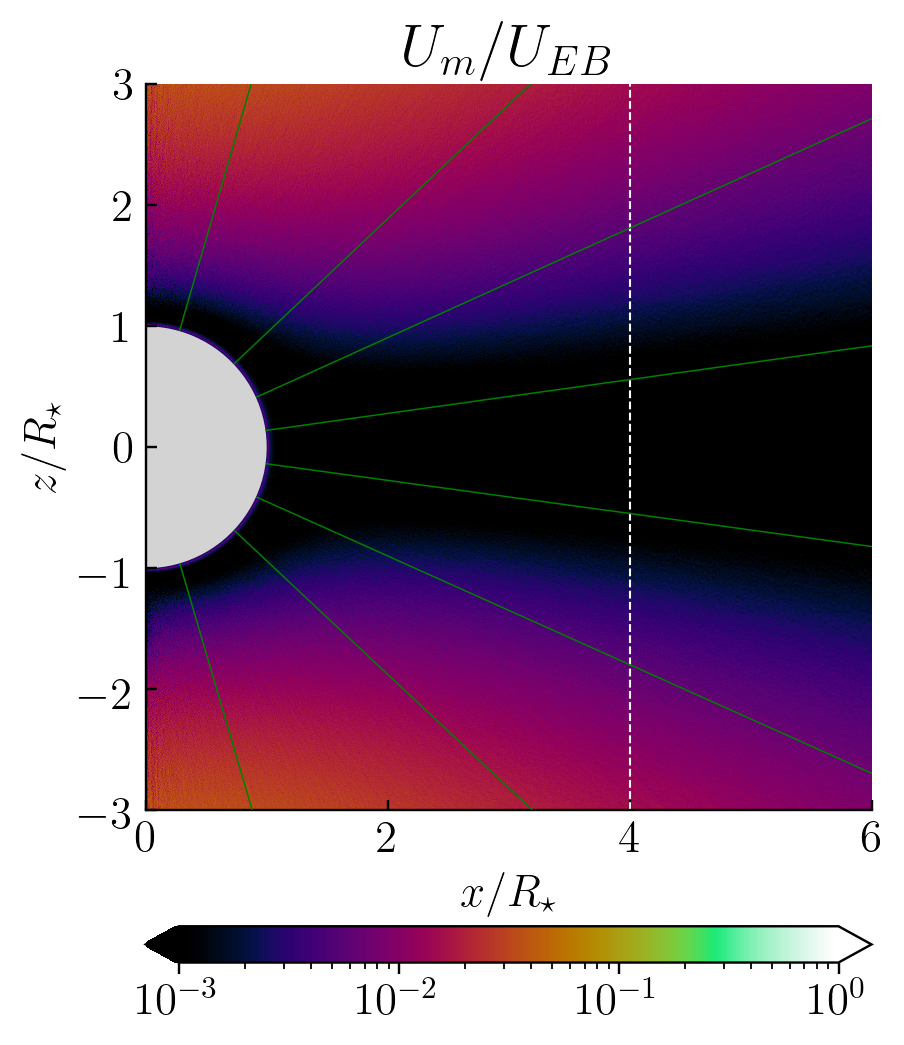}
  \end{tabular}
 \caption{Ratio of the plasma energy density $U_m$ and the electromagnetic
    field energy density $U_{EB} = (E^2+B^2)/(8\pi )$ in Model~I (left) and
    Model~II (right). }
  \label{fig:UmUEB}
\end{figure*}
% %%%%%%%%%%% FIGURE %%%%%%%%%%%%%%%%%%

% %%%%%%%%%%% FIGURE %%%%%%%%%%%%%%%%%%
\begin{figure*}[t!]
    \centering
    \begin{tabular}{cc}
        \includegraphics[width=0.46\textwidth]{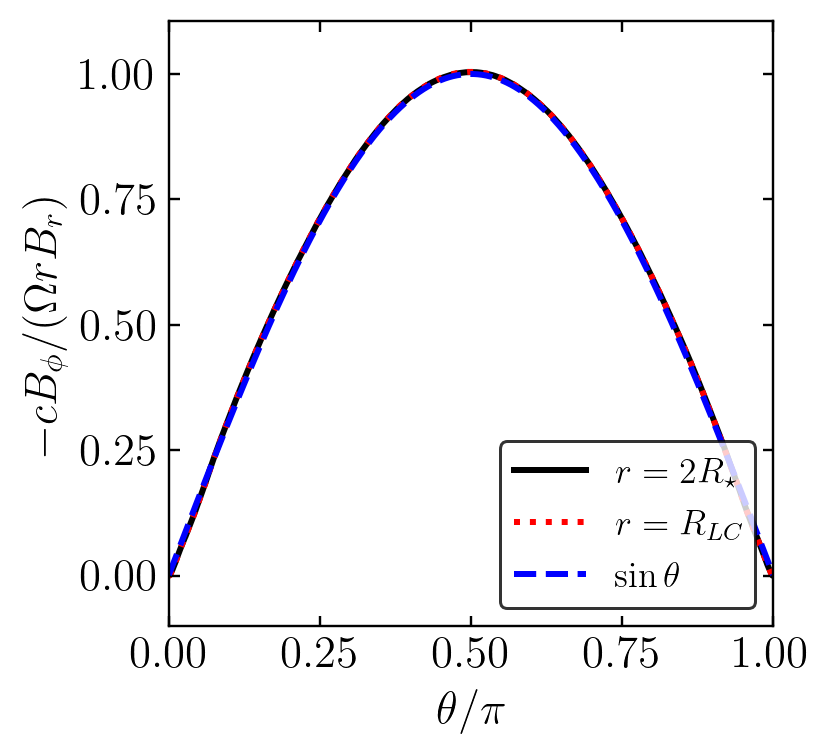}
    &
      \includegraphics[width=0.46\textwidth]{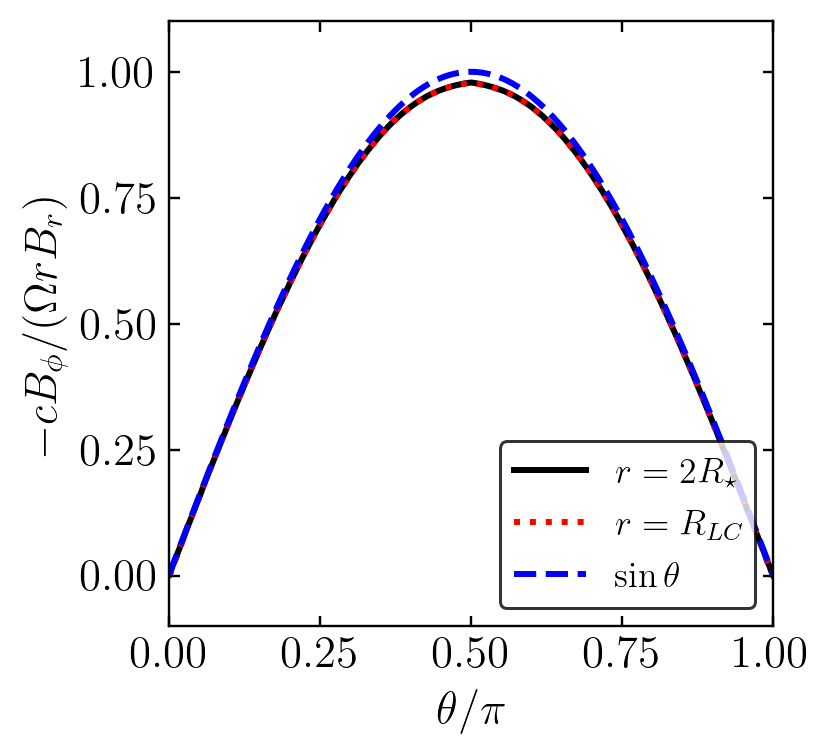}
    \end{tabular}
\caption{Left: Comparison of the electromagnetic configuration obtained in Model~I (MHD wind) with the FFE model, which predicts $-cB_\phi/\Omega r B_r=\sin\theta$ (blue dashed curve). The quantity $-cB_\phi/\Omega r B_r$  measured in the simulation is shown at $r=2R_\star$ (black solid) and at $r=\RLC$ (red dotted). Right: A similar comparison for Model~II (charge separated wind).}
    \label{fig:cs-FF}
\end{figure*}
% %%%%%%%%%%% FIGURE %%%%%%%%%%%%%%%%%%

The wind is magnetically dominated in both Model~I and Model~II (Figure~\ref{fig:UmUEB}), and hence the electromagnetic configuration should be close to Michel's FFE solution given by Equations~(\ref{eq:B_FFE}) and (\ref{eq:E_FFE}). As expected, we observe that the deviations from FFE are small. The poloidal magnetic field is almost exactly radial, and $B_\phi$ closely follows $B_\phi^{\rm FFE}$ given in Equation~(\ref{eq:B_FFE}) (Figure~\ref{fig:cs-FF}). In particular, in Model~II, the deviation $\delta B_\phi/B_\phi^{\rm FFE}\sim \sigma^{-1/2}=2$\%. Note that $\delta B_\phi/B_\phi^{\rm FFE}>0$ in the MHD regime, as the additional inertia due to mass loading increases the bending of magnetic field lines. By contrast, $\delta B_\phi/B_\phi^{\rm FFE}<0$ in Model~II. The nearly force-free electromagnetic configuration enforces the charge density $\rho=\nabla\cdot \bE/4\pi$ and the electric current $\bj=(c/4\pi)\nabla\times\bB$. Therefore, $\rho$ and $\bj$ closely follow the FFE result (Equation~\ref{eq:current}).

% %%%%%%%%%%% FIGURE %%%%%%%%%%%%%%%%%%
\begin{figure*}[t!]
    \centering
        \includegraphics[width=0.46\textwidth]{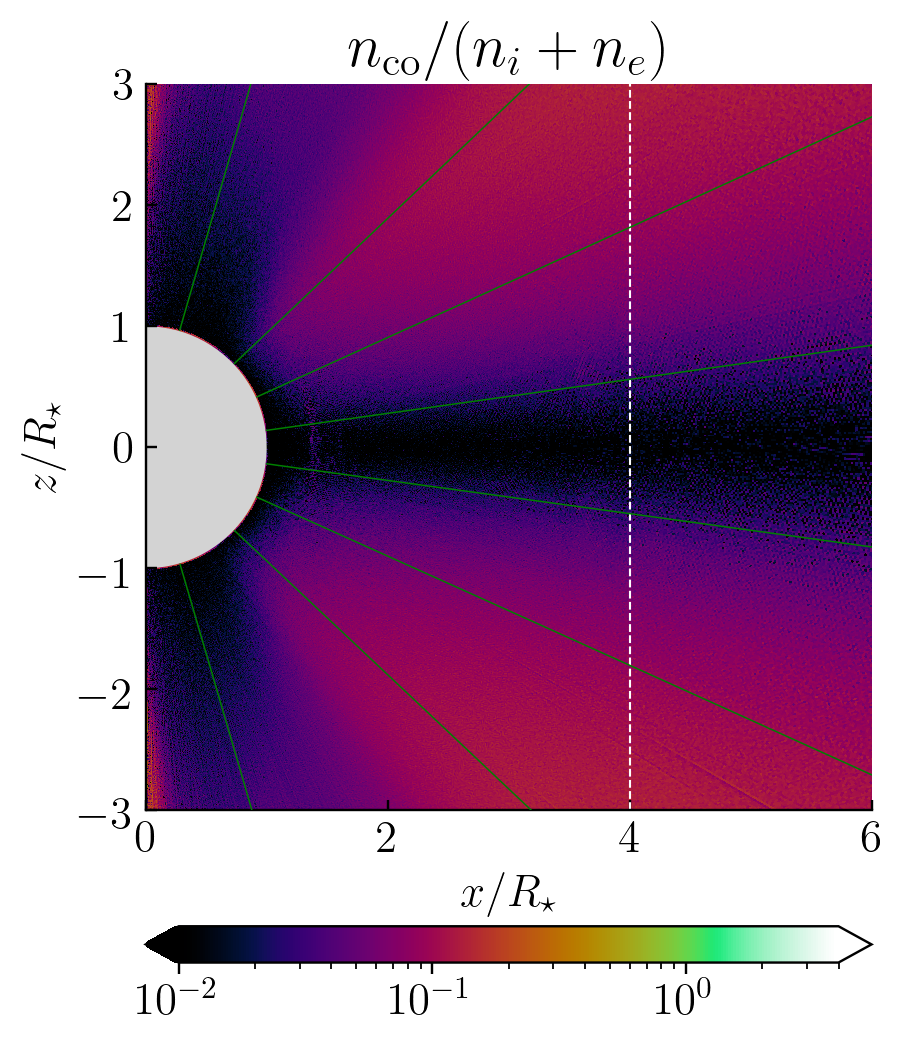}
        \includegraphics[width=0.46\textwidth]{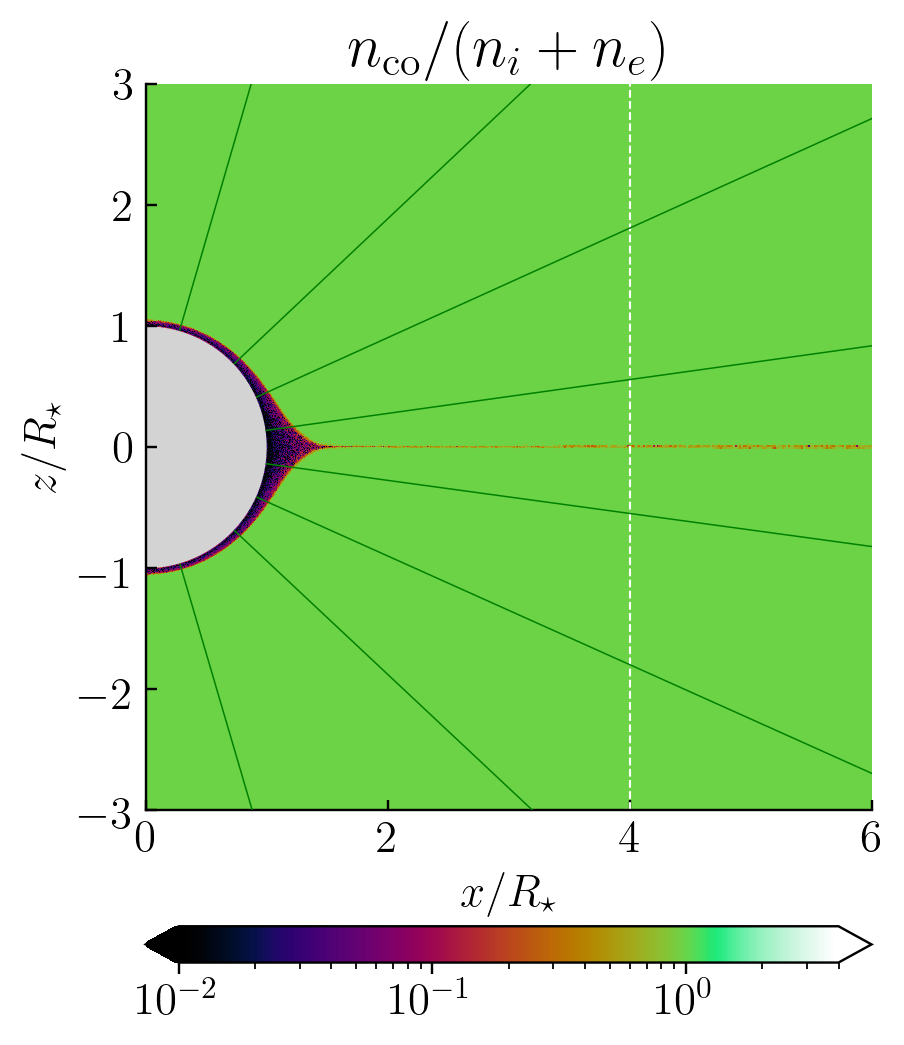}
        \caption{Ratio of the local $n_{co}=\rho_{co}/e$ and the actual particle number density $n = n_i+n_e$ in Model~I (left) and Model~II (right). Model~I has no gravity and the MHD outflow is formed with a high multiplicity $\M$ at all latitudes. Model~II forms a charge-separated outflow ($\M=1$) above the thin gravitationally atmosphere; the outflow occurs in the MHD regime only near the equatorial plane where $j$ and $n_{\rm co}$ vanish.}
    \label{fig:multiplicity}
\end{figure*}
% %%%%%%%%%%% FIGURE %%%%%%%%%%%%%%%%%%

The state of the plasma sustaining this nearly FFE configuration is very different in the charge-separated and MHD regimes. The basic difference between Models~I and II is demonstrated in Figure~\ref{fig:multiplicity}, which compares the plasma density $n=n_++n_-$ with the local $n_{\rm co}$. The MHD regime occurs with abundant supply of $\pm$ charges, $n\gg n_{\rm co}$. Then, the required $\rho\approx\rho_{\rm co}$ and $j\approx c\rhoco$ are achieved  with $|n_+-n_-|\ll n$. In the charge separated regime, $n_+\rightarrow 0$ and the minimum $n=n_-\approx n_{\rm co}$ is established everywhere except the narrow region around the equatorial plane $\theta\approx \pi/2$.

At $\theta=\pi/2$, the monopole wind is not charge separated even in Model~II, simply because $n_{\rm co}=0$ and $j=0$ in the equatorial plane. The surface atmosphere with any finite hydrostatic scale-height  produces a finite outflow  $F\neq  0$, and so, $n/n_{\rm co}\rightarrow \infty$ at $\theta=\pi/2$. As observed in the simulation, Model~II creates an MHD-type outflow in a finite equatorial region $|\theta-\pi/2|<\delta\theta$. The extent of this region $\delta\theta$ may be estimated by evaluating $F_{\rm MHD}$ from Equation~(\ref{eq:F}) (which assumes an MHD outflow with  $E_\parallel=0$) and then comparing $F_{\rm MHD}$ with $|j(r,\theta)|/e$. The boundary of the equatorial MHD wind in Model~II is at the angle $\theta$ where $|j|/e\sim F_{\rm MHD}$.

% %%%%%%%%%%% FIGURE %%%%%%%%%%%%%%%%%%
\begin{figure*}[t!]
    \centering
    \begin{tabular}{cc}
            \includegraphics[width=0.46\textwidth]{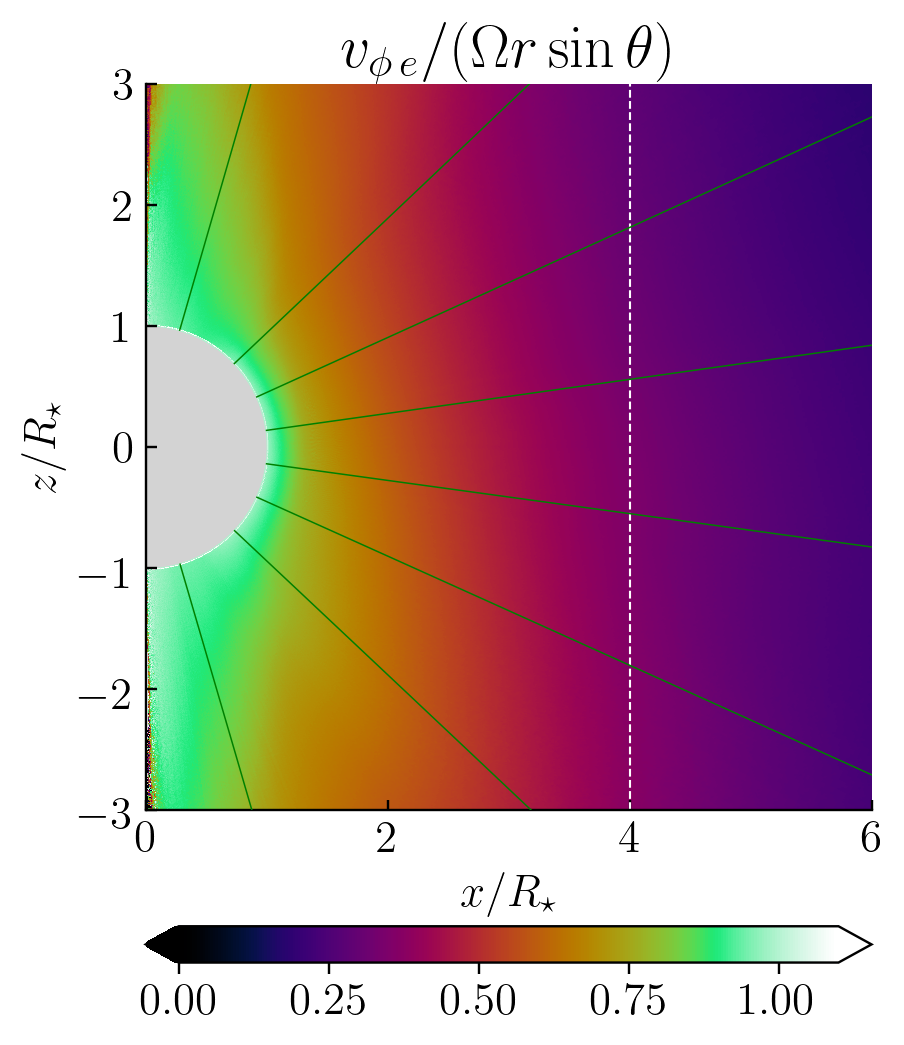} 
        \includegraphics[width=0.46\textwidth]{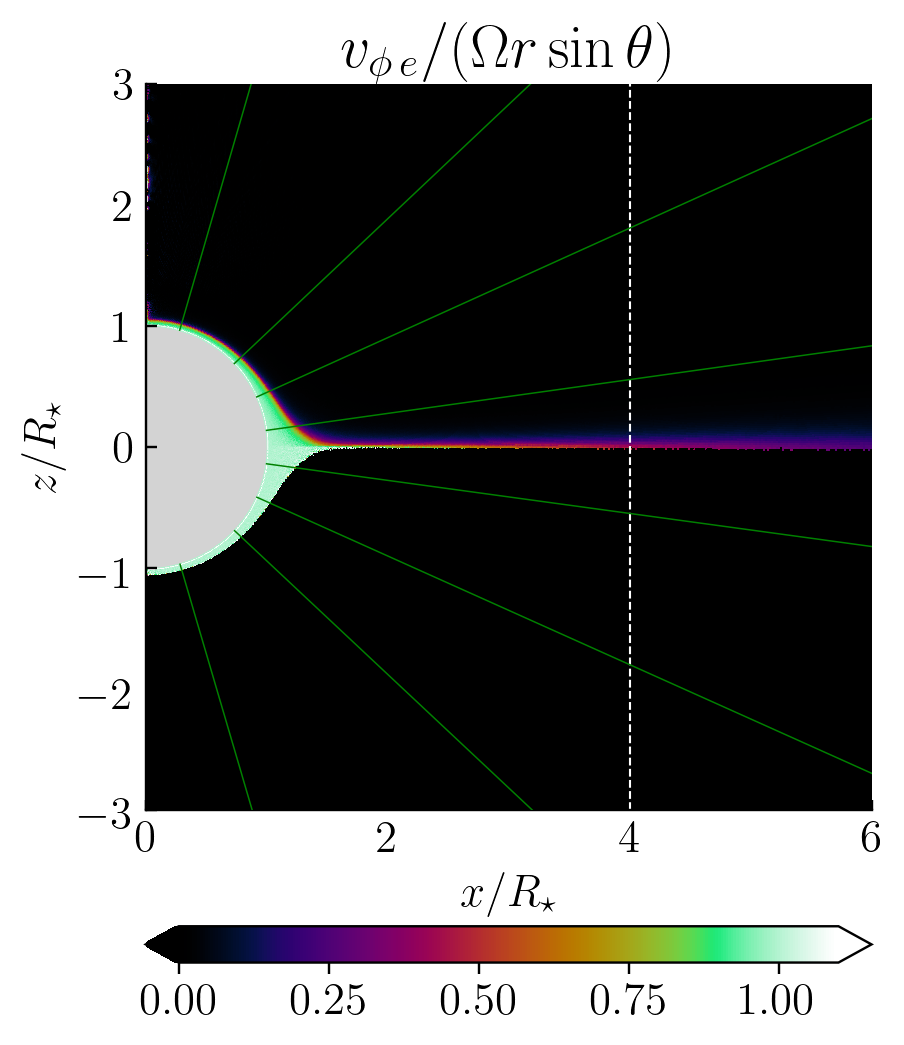}
    \end{tabular}
    \caption{Azimuthal component of electron velocity $v_{\phi}$ for Model~I (left) and Model~II (right).}
    \label{fig:Vphi}
\end{figure*}
% %%%%%%%%%%% FIGURE %%%%%%%%%%%%%%%%%%

In both Model~I and Model~II, the electric current is almost exactly radial, however the bulk motion of the plasma is different in the two models. This is demonstrated in Figure~\ref{fig:Vphi}, which shows the azimuthal component of the plasma bulk velocity. In Model~I, the plasma behaves as a single MHD fluid forced into rotation by the strong magnetic field and flowing out with the centrifugal acceleration. In Model~II the plasma bulk motion is almost exactly radial; the plasma is charge-separated, $n=n_-$, and so the radial current implies the radial bulk motion.

\subsection{MHD wind (Centrifugal Acceleration)}

%%%%%%%%%%%% FIGURE %%%%%%%%%%%%%%%%%%
\begin{figure}[t!]
   \includegraphics[width=0.4\textwidth]{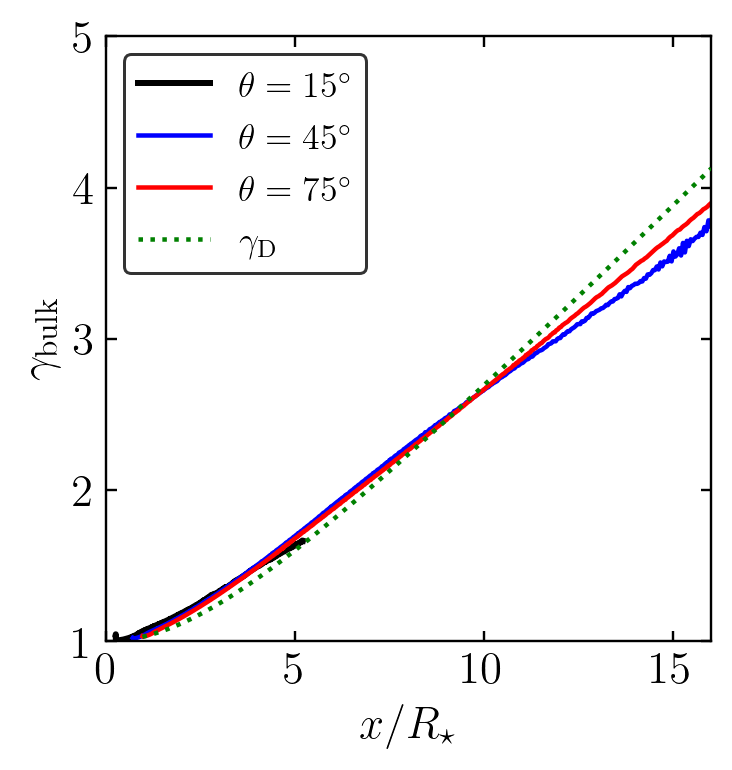}
\caption{Wind Lorentz factor $\gamma$ measured in Model~I (MHD wind). The Lorentz factor is shown as a function of cylindrical radius $x=r\cos\theta$ at a few selected polar angles $\theta$ (solid curves). Green dotted curve shows Lorentz factor $\gamma_{\rm D}$ of the drift motion $\bv_{\rm D}=c\bE\times\bB /B^2$ evaluated using the FFE solution for $\bB$ and $\bE$.}
\label{fig:mhd-radial-profile}
\end{figure}
% %%%%%%%%%%% FIGURE %%%%%%%%%%%%%%%%%%

Figure~\ref{fig:mhd-radial-profile} shows the radial profile of the wind Lorentz factor $\gamma(r)$ measured in Model~I, for a few polar angles $\theta=const$. The plasma bulk motion observed in the simulation is compared with the $\bE\times\bB$ drift. An ideal cold MHD wind is not expected to  develop any motion parallel to the magnetic field lines, because $E_\parallel=0$, so the particles can only have the $\bE\times\bB$ drift motion.  Since $\bE$ and $\bB$ are well described by the FFE solution, the plasma is expected to flow with velocity $\bv\approx\bv_{\rm D}$ given by Equation~(\ref{eq:vD}) and Lorentz factor $\gamma\approx\gamma_{\rm D}$ given by Equation~(\ref{eq:gD}). This is indeed observed in Figure~\ref{fig:mhd-radial-profile}. Small deviations of $\gamma$ from $\gamma_{\rm D}$ of the FFE model are caused by the finite mass loading of the wind and the finite temperature of the plasma injected at the stellar surface. At large cylindrical radii, $\gamma_{\rm D}$ is expected to saturate (Section~\ref{sec:mhd-wind}). This saturation should occur outside the simulation box $R_{\rm out}$ and therefore is not observed in our simulation.

%%%%%%%%%%%%%%%%%%%%%%%%%%%%%%%%%%%%%%
\begin{figure}[h]
   \includegraphics[width=0.4\textwidth]{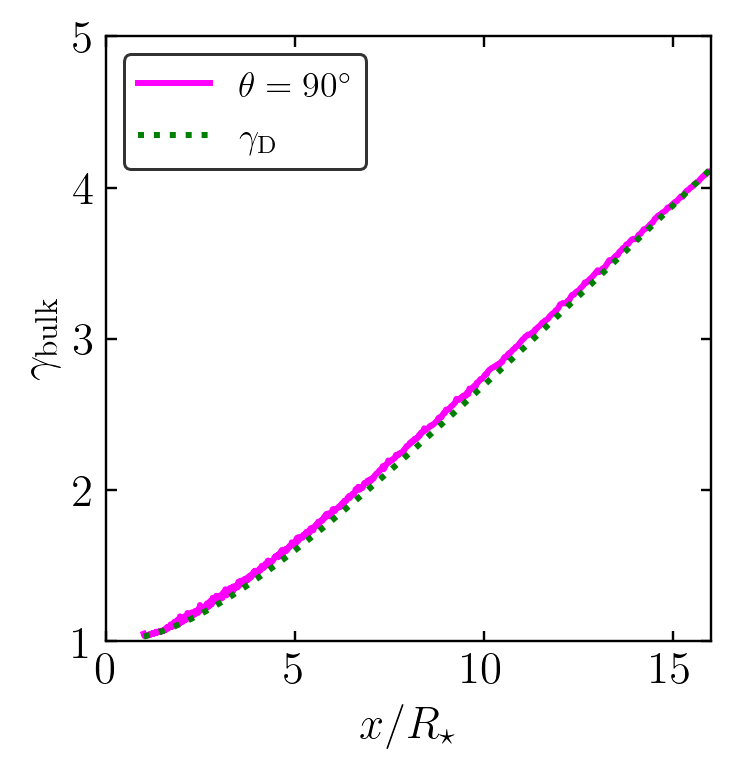}
\caption{Wind Lorentz factor $\gamma$ in Model~II in the equatorial plane $\theta=\pi/2$ where the outflow occurs in the MHD regime (solid curve). Green dotted curve shows Lorentz factor $\gamma_{\rm D}$ of the drift motion $\bv_{\rm D}=c\bE\times\bB /B^2$ evaluated using the FFE solution for $\bB$ and $\bE$.}
\label{fig:cs-radial-profile-eq}
\end{figure}
%%%%%%%%%%%%%%%%%%%%%%%%%%%%%%%%%%%%%%

Figure~\ref{fig:cs-radial-profile-eq} shows the $\gamma(r)$ profile of the equatorial outflow of Model~II. Here, the wind is also in the MHD regime, $n\gg n_{\rm co}$, and we find that the wind moves approximately with the drift Lorentz factor $\gamma_{\rm D}(r)$ of the FFE solution.

% \newpage

\subsection{Charge-separated Wind \\ (Electrostatic Acceleration)}
\label{sec:electr-accel}

% %%%%%%%%%%% FIGURE %%%%%%%%%%%%%%%%%%
\begin{figure*}[t!]
    \begin{tabular}{cc}    
     \includegraphics[width=0.46\textwidth]{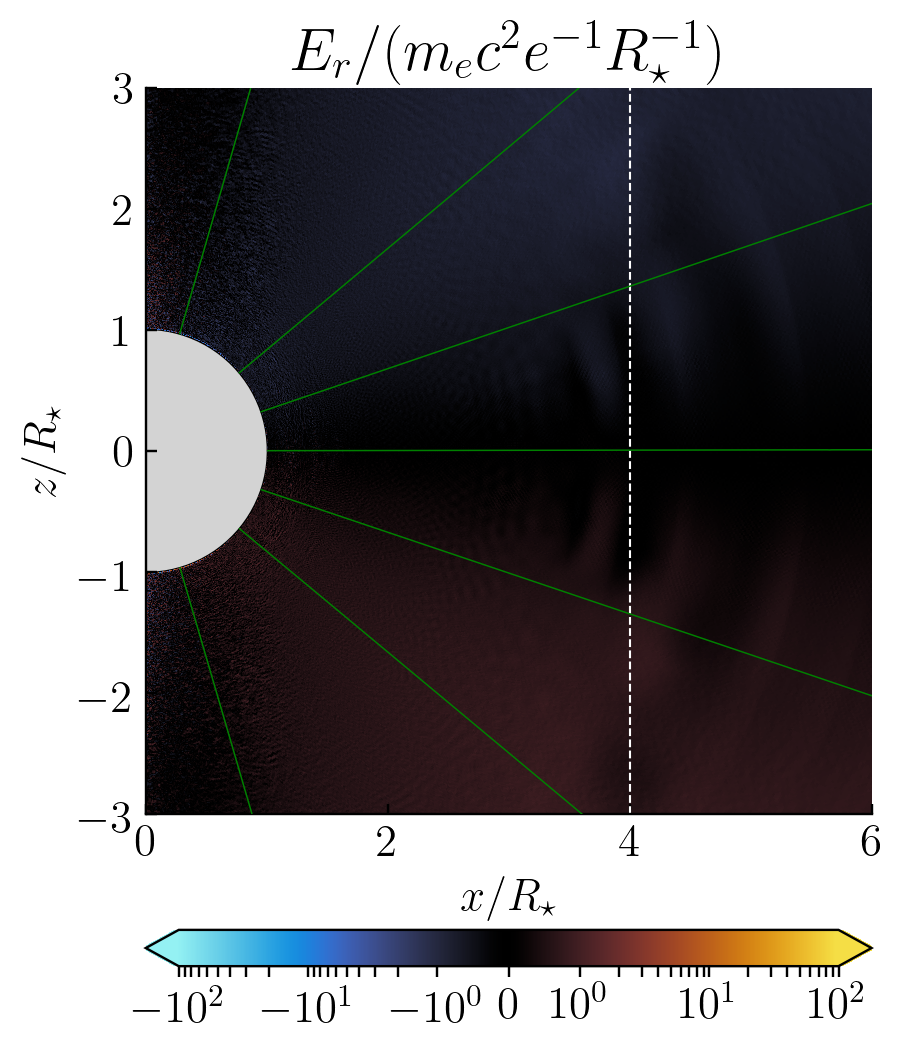} 
        \includegraphics[width=0.46\textwidth]{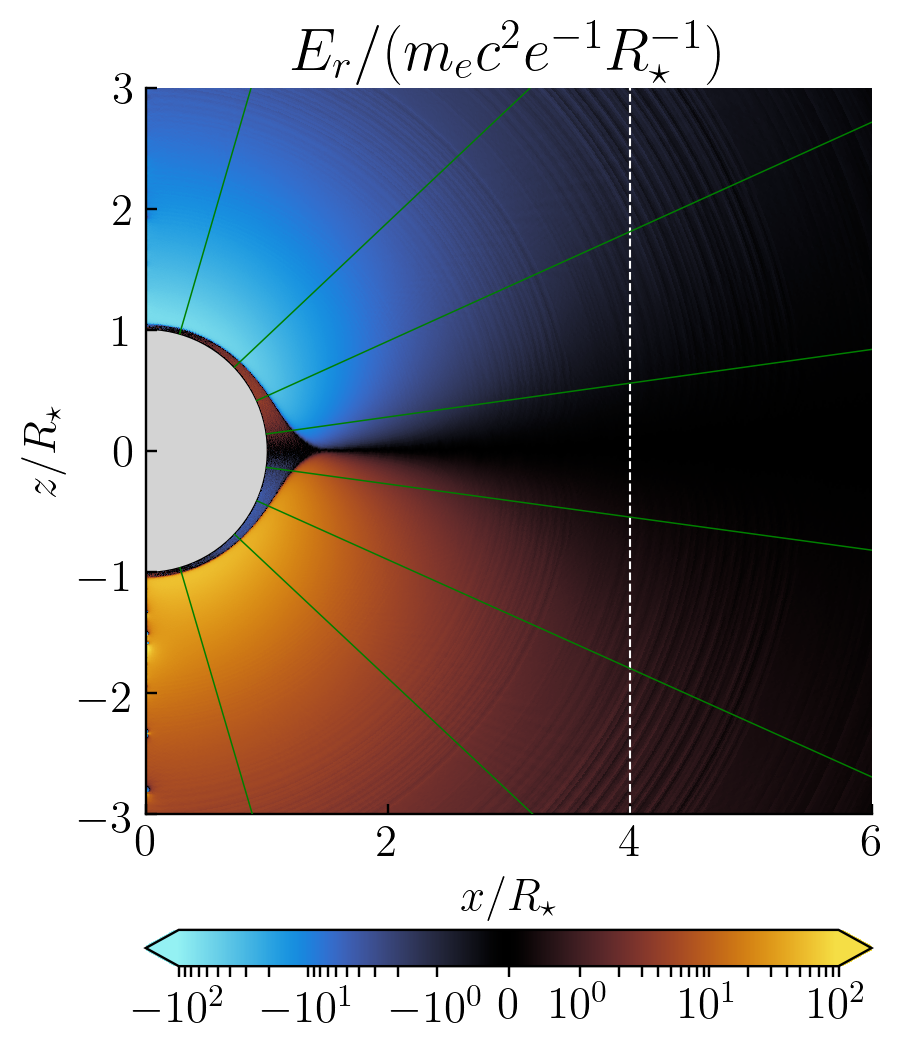}
    \end{tabular}
        \caption{Radial component of the electric field $E_r$ responsible for the wind acceleration in Model~I (left) and Model~II (right).}
    \label{fig:Er}
\end{figure*}
% %%%%%%%%%%% FIGURE %%%%%%%%%%%%%%%%%%

Almost all of the wind in Model~II is charge-separated and accelerated by the electrostatic mechanism: the particles are accelerated away from the star by a strong $E_\parallel=\bE\cdot\bB/B$. Note that $E_\phi=0$ in the steady state and $E_\parallel=E_rB_r/B$. The value of $E_r$ measured in the simulation is shown in Figure~\ref{fig:Er}. It is consistent with the estimate~(\ref{eq:Er}) in Model~II, and greatly exceeds $E_r$ measured in Model~I (also shown in Figure~\ref{fig:Er}, for comparison).  

% %%%%%%%%%%% FIGURE %%%%%%%%%%%%%%%%%%
\begin{figure}[t]
\begin{tabular}{c}
   \includegraphics[width=0.45\textwidth]{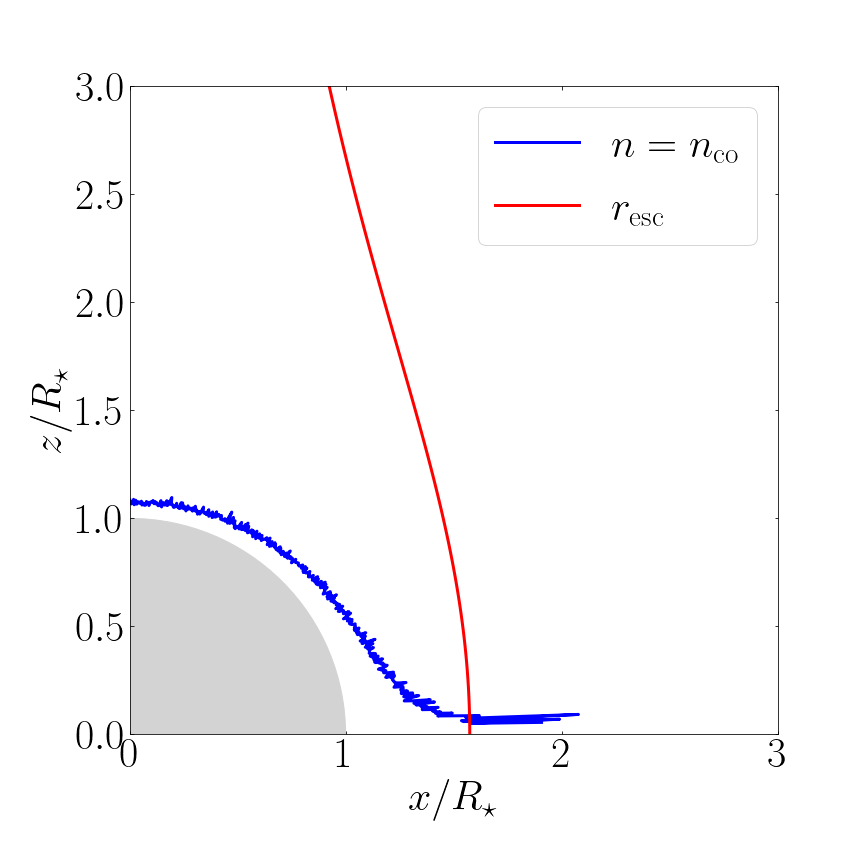}
\end{tabular}
\caption{The surface of wind launching in Model~II (blue curve). At this ``critical'' surface, $r_{\rm cr}(\theta)$, the atmosphere density $n$ drops below $n_{\mathrm{co}} c / v_{\mathrm{th}}$ and a charge-separated wind is extracted by the generated $E_\parallel$. For comparison, the orange curve shows the launching surface $r_{\rm esc}(\theta)$ that would be predicted by the MHD wind model with $E_\parallel=0$ (Equation~\ref{eq:escape-radius}). The MHD regime is realized only where $r_{\rm esc}<r_{\rm cr}$, which occurs near the equatorial plane in Model~II.}
\label{fig:cs-atm-contours}
\end{figure}
% %%%%%%%%%%% FIGURE %%%%%%%%%%%%%%%%%%

We can estimate the expected location of a ``critical'' surface $r_{\rm cr}(\theta)$ outside of which $E_\parallel$ must develop to sustain $j$. 
The critical surface is determined by the condition
\beq
    v_{\rm th}\, n(r_{\rm cr}) = \frac{|j|}{e} \approx c n_{\rm co},
\eeq 
where $n(r)$ is the hydrostatic atmosphere density given by Equation~(\ref{eq:n}).  The solution for $r_{\rm cr}(\theta)$ for Model~II is shown in Figure~\ref{fig:cs-atm-contours}. A charge separated wind is expected to form outside the critical surface, and this is indeed observed in the simulation: the boundary of the region $n\approx n_{\rm co}$ found in Model~II approximately agrees with $r_{\rm cr}(\theta)$. Figure~\ref{fig:cs-atm-contours} also compares $r_{\rm cr}(\theta)$ with $r_{\rm esc}(\theta)$. The equatorial MHD wind occupies the range $\delta\theta$ where $r_{\rm cr}(\theta)>r_{\rm esc}(\theta)$.

\begin{figure}[t]
\begin{tabular}{c}
   \includegraphics[width=0.4\textwidth]{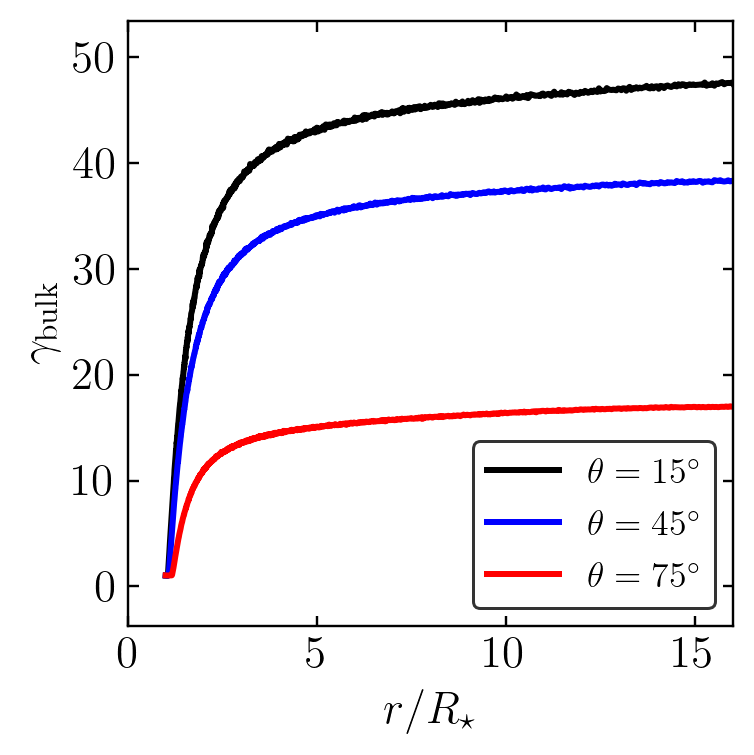}
\end{tabular}
\caption{Radial profile $\gamma(r)$ of the charge-separated wind in Model~II at a few selected polar angles $\theta$.}
\label{fig:cs-radial-profile}
\end{figure}

Figure~\ref{fig:cs-radial-profile} shows the Lorentz factor $\gamma(r,\theta)$ of the electrostatically accelerated wind observed in our simulation for a few selected angles $\theta=const$. As expected from the analytical estimate in Section~\ref{sec:charge-separ-wind}, $\gamma(r)$ rises sharply over a length scale comparable to $R_{\star}$ and then approaches the saturated value $\gamma_{\infty}$.

%%%%%%%%%%%%%%%%%%%%%%%%%%%%%%%%%%%%%%
%%%%%%%%%%%%%%%%%%%%%%%%%%%%%%%%%%%%%%

%%%%%%%%%%%%%%%%%%%%%%%%%%%%%%%%%%%%%%
\begin{figure}[t]
\begin{tabular}{c}
   \includegraphics[width=0.4\textwidth]{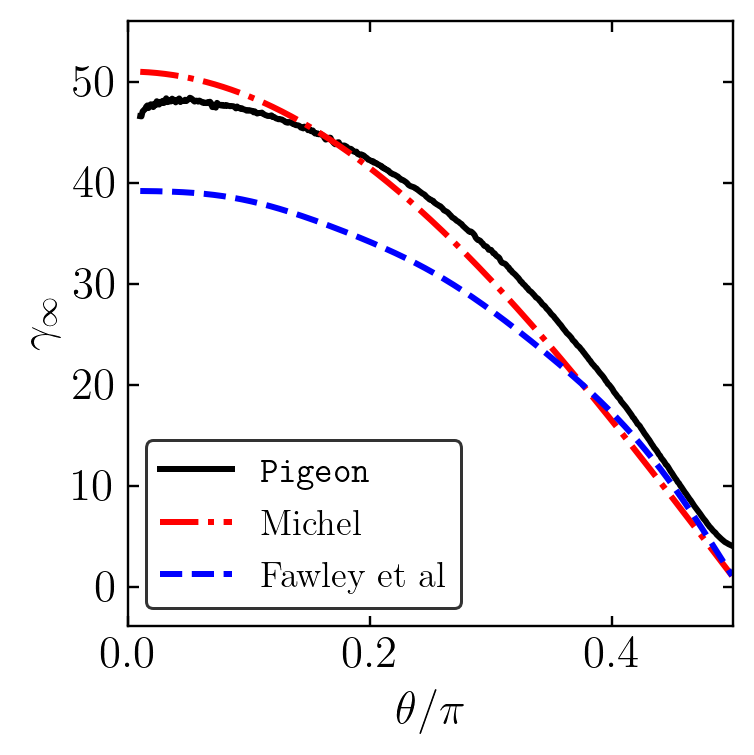}
\end{tabular}
\caption{Comparison of $\gamma_{\infty}(\theta)$ given by different models.
Michel's (1974) approximation is shown by the red dash-dotted curve, the result of Fawley et al. (1977) is shown by the blue dashed curve, and our simulation is shown by the black solid curve.}
\label{fig:cs-gamma-final}
\end{figure}
%%%%%%%%%%%%%%%%%%%%%%%%%%%%%%%%%%%%%%

The dependence of the asymptotic $\gamma_{\infty}$ at $r\rightarrow \infty$  on the polar angle $\theta$ is shown in Figure~\ref{fig:cs-gamma-final}. It was measured at a finite, but sufficiently large radius $r=15R_\star\gg R_\star$. For comparison, we also show the solutions of \citet{michel_rotating_1974} and \citet{fawley_potential_1977} discussed in Section~\ref{sec:charge-separ-wind} (Equations~\eqref{eq:Michel} and \eqref{eq:Fawley}). The Legendre series in Equation~\eqref{eq:Fawley} were truncated at $l=5$. Michel's approximation appears to agree better with our simulation results.

%###########################################

\section{Discussion}

In this paper, we have investigated the relativistic plasma wind from a rotating star with a monopole magnetic field. The star was assumed to have a thermal atmosphere, which supplied plasma to the wind. This provides a complete setup for the self-consistent global study of the wind formation and acceleration driven by the rotation of the star. We have used the monopolar setup as a first test problem for our PIC code \texttt{Pigeon} developed for global kinetic plasma simulations of the plasma behavior around rotating compact objects. 

Our simulations have demonstrated the wind formation in two qualitatively different regimes. The first regime gives a centrifugally accelerated, nearly ideal ($E_\parallel=0$), MHD wind (Model~I). It occurs if the thermal atmosphere of the star supplies enough plasma  at high altitudes to form an outflow with density $n\gg n_{\rm co}=|\boldsymbol{\Omega}\cdot\bB|/2\pi  c e$. This condition is violated when the atmosphere on the stellar surface is sufficiently thin. Then, we observed the onset of the second regime: a strong $E_\parallel$ is induced, which extracts charges of one sign from the atmosphere and accelerates them away from the star. As a result, an electrostatically accelerated,  charge-separated wind forms (Model~II). The charge-separated wind develops Lorentz factors $\gamma\sim \sigma^{1/2}$ with the highest $\gamma$ near the polar axis. This electrostatic acceleration is much more powerful than the centrifugal acceleration of the MHD wind. It occurs near the star (rather than beyond the light cylinder), in agreement with the approximate analytical results of \citet{michel_rotating_1974} and \citet{fawley_potential_1977}. 

In general, when electrons develop sufficiently high Lorentz factors $\gamma$, $e^\pm$ creation may be expected. A standard channel for $e^\pm$ creation around neutron stars involves the emission of curvature gamma-rays which convert to $e^\pm$ pairs. However, this mechanism is not activated in the monopole charge-separated wind, even in the limit of arbitrarily high $\sigma$ (which can give any high $\gamma$), because the particles flow out radially with no curvature in their trajectories. 

The rotating monopole magnetosphere is also special because its four-current $j^\mu=(c\rhoco,\bj)$ is null in the magnetically dominated (FFE) limit. Then the parameter $\alpha\equiv j/c\rhoco=1$ is right at the boundary between two distinct regimes: $\alpha<1$ would give and oscillating low-energy outflow, and $\alpha>1$ would give a stronger acceleration \citep{beloborodov_polar-cap_2008}. The special case of $\alpha=1$ can serve as a good test for PIC codes, because it forms a steady ultra-relativistic wind while the fields remain close to the known analytical FFE solution. Real magnetospheres, however, do not have the special $\alpha=1$, as their magnetic field lines are curved. In addition, general relativistic effects (in particular, the frame-dragging effect) in the gravitational field modify $\bj=(c/4\pi)\nabla\times\bB$ and $\rho=\nabla\cdot\bE/4\pi$, and change $\alpha$ from unity \citep{muslimov_general_1992,philippov_ab_2015,Carrasco18}.

Rotating magnetized neutron stars have enormous $\sigma\sim 2\times 10^{12}\,B_{12}\Omega_2$ (Equation~\ref{eq:sigma}), which makes them extremely efficient accelerators. Their magnetospheres are not monopolar, however the idealized monopole model captures some basic features of plasma winds. In particular, the MHD and charge-separated regimes can occur in real objects, and the transition between the two regimes is similar to that in the monopole model. For instance, a nascent hot neutron star is expected to produce an MHD wind during a short period after its birth \citep[e.g.][]{Metzger07}. As the neutron star cools, the plasma supply to the wind drops, and the star becomes capable of generating a charge-separated outflow accelerated to high energies. Then, the neutron star becomes an active pulsar. 

The pulsar magnetosphere has both closed and open magnetic field lines. The outflow along open field lines partially resembles the monopole wind and may be fed by charges extracted by $E_\parallel$ from the star or from its atmospheric layer. The pulsar magnetosphere, however, includes the quasi-dipole magnetic field inside the light cylinder, and this leads to a continual $e^\pm$ discharge. The application of \texttt{Pigeon} to this problem will be presented in the next paper (Hu \& Beloborodov, in preparation).

\medskip

A.M.B. is supported by NASA grant NNX\,17AK37G, NSF grant AST\,2009453, Simons Foundation grant \#446228, and the Humboldt Foundation. A.C. is supported by NSF grants AST-1806084 and AST-1903335.

% \nocite{*}
\bibliographystyle{aasjournal}
\bibliography{ms}

\end{document}